\documentstyle[11pt,aaspp4,epsfig]{article}

\def\simg{\mathrel{\hbox{\rlap{\lower.55ex \hbox {$\sim$}}
                   \kern-.3em \raise.4ex \hbox{$>$}}}}
\def\siml{\mathrel{\hbox{\rlap{\lower.55ex \hbox {$\sim$}}
                   \kern-.3em \raise.4ex \hbox{$<$}}}}
\def\Mesz{M\'esz\'aros~}
\def\Pacz{Paczy\'nski~}
\def\beq{\begin{equation}}
\def\enq{\end{equation}}
\def\bea{\begin{eqnarray}}
\def\ena{\end{eqnarray}}
\def\bec{\begin{center}}
\def\enc{\end{center}}
\def\etal{{\it et al.}}
\def\epm{e^\pm}
\def\cl{{\ell'}}
\def\L52{L_{52}}
\def\Lo{L_o}
\def\Lpht{L_{pht}}
\def\Lk{L_k}
\def\Lsh{L_{sh}}
\def\m1{\mu_1}
\def\ro{r_o}
\def\rp{r_p}
\def\rs{r_s}
\def\rsh{r_{sh}}
\def\rph{r_{ph}}
\def\to{t_o}
\def\tv{t_v}
\def\Go{\Gamma_o}
\def\Gp{\Gamma_p}
\def\Gph{\Gamma_{ph}}
\def\ets{\eta_\ast}
\def\etp{\eta_p}
\def\et2{\eta_2}
\def\Tho{\Theta_o}
\def\Thp{\Theta_p}
\def\Thph{\Theta_{ph}}
\def\taus{\tau_s}
\def\epsh{\varepsilon_{sh}}
\def\esh1{\varepsilon_{sh-1}}
\def\epb{\epsilon_B}
\def\eep{\epsilon_e}
\def\ergs{\hbox{erg s$^{-1}$}}
\def\st{\sigma_T}

\def\msun{M_\odot}
\def\Mbh{M_{bh}}
\def\eps{\epsilon}

\begin{document}


\slugcomment{ApJ'00, in press, astro-ph/9908126} 
\bigskip

\title{ Steep Slopes and Preferred Breaks in GRB Spectra:\\
        the Role of  Photospheres and Comptonization} 

\author{P. \Mesz$^{1,2}$ \& M.J. Rees$^3$ }
\smallskip\noindent
$^1${Dpt. of Astronomy \& Astrophysics, Pennsylvania State University,
University Park, PA 16802}\\
$^2${Institute of Theoretical Physics, University of California, Santa Barbara,
CA 93106-4030}\\
$^3${Institute of Astronomy, University of Cambridge, Madingley Road, Cambridge
CB3 0HA, U.K.}


\begin{abstract}
The role of a photospheric component and of pair breakdown is examined in
the internal shock model of gamma-ray bursts. We discuss some of the mechanisms 
by which they would produce anomalously steep low energy slopes, X-ray excesses
and preferred energy breaks. Sub-relativistic comptonization should dominate
in high comoving luminosity bursts with high baryon load, while synchrotron 
radiation dominates the power law component in bursts which have lower comoving
luminosity or have moderate to low baryon loads. A photosphere leading to 
steep low energy spectral slopes should be prominent in the lowest baryon load 
cases.

\end{abstract}

\keywords{Gamma-rays: Bursts -  Radiation Mechanisms- Cosmology: Miscellaneous}

\newpage

\section{Introduction}

The standard fireball shock scenario for gamma ray bursts (GRB) assumes a 
synchrotron and/or an inverse Compton (IC) spectrum, in good general agreement 
with a number of observations (Tavani 1996; Cohen \etal 1997, Panaitescu, Spada \& 
\Mesz 1999). 
The physical motivation for this scenario is strong; it would indeed be surprising 
if the expansion of the ejecta from the huge energies inferred in GRB did not lead 
to shocks, where synchrotron and IC radiation play a significant role. One need 
only remind oneself of the dominant role of these effects in AGN jets and 
supernova remnants.  However, in a significant fraction of bursts there is 
evidence in the 1-10 keV range for spectral intensity slopes steeper than 1/3 
(photon number slopes flatter than -2/3),  as well as in some cases for a soft 
X-ray excess over the extrapolated power law from higher energies (Preece \etal 
1998, 1999; Crider \etal 1997, etc.), 
and this has served as the motivation for considering a thermal (e.g. Liang (1997) 
or a thermal/nonthermal (Liang \etal 1999) comptonization mechanism. 
While an astrophysical model where this mechanism would arise naturally has been  
left largely unspecified, Ghisellini \& Celotti 1999 have pointed out 
that if internal shocks lead to high compactness parameters and pair formation, 
this could naturally produce conditions where the pair temperature and the 
scattering opacity are self-regulated at values favoring comptonization. 

There is also a trend in current analyses of GRB spectra indicating that the
apparent clustering of spectral break energies in the 50 keV-1 MeV range
is not due to observational selection effects (e.g. Preece \etal 1998; 
Brainerd \etal 1998; see however Dermer \etal 1999a).
Models to explain a preferred break physically, e.g. through a Compton
attenuation model (Brainerd \etal 1998)
require reprocessing by an external medium whose column density adjusts itself
near a few g cm$^{-2}$.
More recently a preferred break has been attributed to the blackbody peak of
the fireball photosphere when this occurs at the comoving pair recombination 
temperature in the accelerating regime, which is blueshifted to the appropriate 
observer frame energy 
(Eichler \& Levinson 1999), and in this case the Rayleigh-Jeans portion of the 
photosphere provides a steep low energy spectral slope. To get a photosphere as 
close in as to occur at the pair annihilation temperature requires an extremely 
low baryon load outflow; the presence of a high energy power law extending to GeV 
then requires a separate explanation.  

At the other extreme of large baryon load outflows, Thompson (1994)
has considered scattering off MHD turbulence in photospheres in the coasting 
regime, which boosts the adiabatically cooled thermal photons near the 
photospheric peak up to a larger break energy, leading to canonical energy slopes 
of 0 and -1 (photon slopes -1 and -2) below and above the break. An alternative 
scenario is that of Ghisellini \& Celotti (1999), 
who invoke $\epm$ pair breakdown in high compactness shocks above the 
photosphere to achieve a self-regulating nonrelativistic lepton temperature 
and scattering optical depth of order a few, leading to a thermal comptonization
spectrum with standardized features. While this does not explain steep low energy
slopes it does provide nonthermal spectral energy power laws of slope $\sim 0$ 
close to those of synchrotron, as well as a standard temperature which would 
lead to a preferred break in the  observer frame, if the bulk Lorentz factor 
is in a narrow range.

In this paper we synthesize and extend some of these ideas within the framework 
of the standard fireball internal shock model. While in our previous work 
(e.g. \Mesz, Laguna \& Rees 1993, Rees \& \Mesz 1994) we considered photospheres 
and pair formation, their thermal character, the strong uncompensated photosphere 
redshift in the coasting phase, and the lack of a straightforward way to get from 
these a power law extending to GeV energies served as compelling arguments for 
concentrating on the synchrotron and inverse Compton mechanisms. 
Here we re-examine the relative roles of 
photospheres and shocks, as well as those of synchrotron, pair breakdown, 
scattering on MHD waves and comptonization. This leads to a unified picture 
where both shocks and/or a photosphere with a nonthermal component can provide 
most of the luminosity, and where the synchrotron and IC mechanisms in shocks 
provide the primary spectrum or (in high comoving luminosity cases) the trigger for 
pair breakdown leading to comptonization. We investigate the range of burst model 
parameters over which different mechanisms come to the fore, and discuss their 
role in providing flatter or steeper spectral slopes as well as preferred 
spectral break energies.

\section{ Fireball Photospheric Luminosity }
\label{sec:phot}

We assume a fireball wind of total luminosity output $\Lo =10^{52}\L52 \ergs$ 
expanding from some initial radius $\ro$, which for the sake of argument is
normalized to the last stable orbit $\ro$ at three Schwarzschild radii around 
a non-rotating black hole of mass $\Mbh = 10 \mu_1 \msun$, with a corresponding 
Kepler rotation timescale $\to$,
\bea
\ro = & 6 G \Mbh /c^2= 0.9\times 10^7 \mu_1 ~\hbox{cm}~,~~~~~~~~~~~ \nonumber \\
\to = & {2\pi \ro^{3/2} (2G\Mbh)^{-1/2}}= 3.25\times 10^{-3}\m1 ~\hbox{s}~.
\label{eq:ro}
\ena
The initial blackbody temperature in 
electron rest mass units at that radius is
\beq
\Tho = (k /m_e c^2 )(\Lo/4\pi \ro^2 c \Go^2 a )^{1/4}= 
      2.42 \L52^{1/4}\mu_1^{-1/2} \Go^{-1/2}~,
\label{eq:Tho}
\enq
(or $\sim$ 1.2 MeV) for a $10\mu_1$ solar mass BH and an initial bulk Lorentz 
factor $\Go \geq 1$ at $\ro$. As the optically thick (adiabatic) wind expands with 
comoving internal energy $\eps' \propto n'^{4/3}$, where $n'$ is the comoving
baryon number density, the baryon bulk Lorentz factor increases as $\Gamma 
\propto r$ and the comoving temperature drops $\Theta' \propto r^{-1}$. 
The $\epm$ pairs drop out of equilibrium (\Pacz 1986, 1990) 
at $\Thp' \sim 0.03$ ($\sim$ 17 keV), at a radius $\rp$ where the bulk Lorentz
factor has grown linearly to a value $\Gp$,
\beq
{\rp \over \ro}= {\Gp \over \Go}= {\Tho \over \Thp}= 
               7\times 10^1 \L52^{1/4} \m1^{-1/2} \Go^{-1/2}~.
\label{eq:rp}
\enq
This is the radius of an $\epm$ pair photosphere, above which the scattering 
optical depth is less than unity, unless the wind carries enough baryons to
provide an electron scattering photosphere above $r_p$. For a wind baryon load 
$\dot M$ parameterized by a dimensionless entropy $\eta=L/ {\dot M} c^2 $, the 
baryonic electrons lead to a photosphere larger than equation (\ref{eq:rp}) if 
$\eta < \eta_p$ (equation [\ref{eq:etp}]). As long as the wind remains optically 
thick it is radiation dominated and continues to expand as a relativistic 
gas with $\Gamma \propto r$ (e.g. Shemi \& Piran 1990). 
Clearly $\Gamma$ cannot exceed $\eta=\Lo/{\dot M} c^2$, and for large baryon loads 
or moderate $\eta$ this occurs at a saturation radius $\rs/\ro= \eta/\Go$ 
(for low loads or large $\eta$ see however equation [\ref{eq:Gammas}]).
Above the saturation radius, the flow continues to coast with $\Gamma=$ constant
equal to the final value achieved at $\rs$.

An electron scattering photosphere is defined by $\taus'=n' Y \st\rph /\Gamma =1$, 
where $n'=(L/4\pi r^2 m_p c^3 \Gamma \eta)$ is the comoving baryon density, $Y$ is 
the number of electrons per baryon and $r/\Gamma$ is a typical comoving length. 
For relatively low values of $\eta < \ets$ (defined in equation [\ref{eq:ets}])
the flow remains optically thick above the saturation radius $\rs$ and the 
photosphere arises in the coasting regime $\Gamma =\eta=$ constant, at a radius 
$\rph >\rs$, (Rees \& \Mesz, 1994, Thompson 1994), 
\bea
{\rph^> \over \ro}= & {L \st Y \over 4\pi \ro m_p c^3 \eta^3}=
 1.3\times 10^6 \L52 \m1^{-1} Y \eta_2^{-3} \nonumber \\
                  = & \Go^{-1} \ets (\eta/\ets)^{-3}~~~~~~~~~~~~~~~~~~~~~~~~~~.
\label{eq:rph>}
\ena
Here $\ets$ is the critical value at which $\rph^> =\rs$,
\beq
\ets= \left( { L \st Y \Go \over 4\pi m_p c^3 \ro} \right)^{1/4} \simeq
 10^3 (\L52 \m1^{-1} Y \Go)^{1/4}~,
\label{eq:ets}
\enq
which is the wind equivalent of the critical $\eta$ of an impulsive fireball
photosphere discussed in \Mesz, Laguna \& Rees 1993, labeled there $\eta_m$. 

For low baryon loads where $\eta > \ets$ a baryonic electron photosphere appears
in the accelerating portion $\Gamma \propto r$ of the flow at $\rph <\rs$, 
\bea
{\rph^< \over \ro}= & \left( {L\st Y \over 4\pi \ro m_p c^3 \eta \Go^2 } \right)^{1/3} =
   2.35\times 10^3 (\L52 Y \Go^{-2})^{1/3} \m1^{-1/3} \et2^{-1/3}\nonumber \\
                  = & \Go^{-1} \ets (\eta/\ets)^{-1/3}~~~~~~~~~~~~~~~~~~~~.
\label{eq:rph<}
\ena
These photospheric radii are shown in Figure 1.
For sufficiently high $\eta$, the baryon photospheric radius given by equation 
(\ref{eq:rph<}) can formally become smaller than the pair photosphere radius of 
equation (\ref{eq:rp}), in which case the latter should be used. The minimum 
possible photospheric radius is therefore achieved at $r=\rp$ given by equation
(\ref{eq:rp}), requiring extremely large values of $\eta >\etp $ ($>\ets$),
\beq
\etp= 4\times 10^6 \L52^{1/4} \m1^{1/2} Y \Go^{-1/2}=
       3.75 \times 10^3 \ets (\m1 Y \Go^{-1})^{3/4}~,
\label{eq:etp}
\enq
which implies extremely low baryon loads, ${\dot M} \leq 1.5\times 10^{-8}
\msun {\rm s}^{-1}$.

The Lorentz factor attained at the photosphere, $\Gamma_{ph}$, grows linearly 
with $\eta$ for $1 \leq \eta \leq \ets$ (assuming $\Go=1$), then it decays as 
$\eta^{-1/3}$ up to $\etp$ and remains constant above that.
The comoving temperature $\Thph' \propto n'^{1/3} \propto r^{-1}$ for a
photosphere at $\rph <\rs$, while $\Thp' \propto r^{-2/3}$ for $\rph >\rs$. The 
observer-frame photospheric temperature $\Thph = \Thph' \Gph$ is then
\beq 
{\Thph \over \Tho} = \cases{ 
   (\rph/\rs)^{-2/3}= (\eta/\ets)^{8/3},& for $\eta <\ets,~\rph>\rs$; \cr
                  1                    ,& for $\eta >\ets,~\rph <\rs$. \cr}
\label{eq:Thph}
\enq
The observed photospheric thermal luminosity is $\Lpht \propto
r^2 \Gamma^2 \Thph'^4 \propto r^0$ for $r<\rs$ and $\Lpht \propto r^{-2/3}$ for
$r>\rs$, hence
\beq
{\Lpht \over \Lo} = \cases{ 
         (\rph^> /\rs)^{-2/3}=(\eta/\ets)^{8/3},
                                         & for $\eta <\ets,~\rph>\rs$;\cr 
                           1 ,           & for $\eta>\ets,~\rph<\rs$.\cr}
\label{eq:Lpht}
\enq

\section{ Kinetic and Internal Shock Luminosity}
\label{sec:kin}

For typical values of $\eta \siml 10^3$ it is clear that the terminal 
baryon Lorentz factor is $\Gamma=\eta$, since the photosphere occurs beyond the 
saturation radius after the baryons are already coasting with $\eta$. However the 
terminal baryon Lorentz factor is less obvious in cases where $\eta >\ets$.  For 
such values, a photosphere occurs in the regime where $\Gamma \propto r$, so 
$\rph<\rs$, but the question is what happens to the baryons above this 
photosphere, and what is the appropriate value of $\rs$. 
One possibility is that the outflow has magnetic fields strong enough that
Poynting stresses continue to accelerate baryons outside the photosphere. If
radiation provides the dominant relativistic pressure, to achieve a saturation
radius at the value $\rs/\ro=\eta/\Go$ would require the baryons to be coupled to
the radiation out to that radius, beyond  the photosphere. Alternatively, it 
is sometimes assumed that the baryons decouple at the photosphere $\rph$, and
coast thereafter with $\Gamma=\Gamma_{ph}=\Go (\rph/\ro)$. However, the fact that 
the outflow has become optically thin to scattering means that most photons no 
longer scatter. Nonetheless, most of the electrons above the photosphere can still
scatter with a decreasing fraction of free-streaming photons, as long as 
the comoving Compton drag time $t'_{dr}=m_p c^2/c\st u'_\Gamma$ is less than the 
comoving expansion time $t'_{ex}=r/c\Gamma$. The ratio of these two times, 
\beq
(t'_{dr}/t'_{ex})=(4\pi m_p c^3 r \Gamma^3/L\st)=(\ets/\Go)^{-4}(r/\ro)^4~,
\label{eq:tdrag}
\enq
exceeds unity above a radius $r_\ast/\ro= \ets / \Go$. Thus for $\eta>\ets$
the appropriate saturation Lorentz factor is $\ets$ (instead of the larger $\eta$) 
and the saturation radius is $\rs/\ro=r_\ast/\ro=\ets /\Go < \eta/\Go$. Thus, 
in general the terminal bulk Lorentz factor and the saturation radius are
\beq
\Gamma_s=\min ~[\eta,\ets]~~~,~~~
(\rs / \ro )= \min ~[\eta,\ets] / \Go ~,
\label{eq:Gammas}
\enq
where the critical value $\ets$ is given by equation (\ref{eq:ets}). 

The observer-frame kinetic (matter) luminosity of the outflow $\Lk \propto
r^2 \Gamma^2 n' (kT' + m_p c^2) \propto r$ for $r<\rs$ and $\Lk \propto r^0$ for
$r >\rs$. For $\eta <\ets$ it is clear that $\Lk$ reaches
the level $\Lo = {\dot M} c^2 \eta$ since the terminal bulk Lorentz factor 
saturates at the initial dimensionless entropy $\eta$. However, for 
$\eta >\ets$ the terminal $\Lk$ can only reach a lower level, since the bulk 
Lorentz factor saturates at the lower value $\ets <\eta$. The terminal value
of $\Lk$ above $\rs$ is then
\beq
{\Lk \over \Lo}= \cases{ 
                   1 ~,                 &~ for  $\eta<\ets,~r>\rs$; \cr
                  (\ets/\eta) \leq 1 ~, &~ for $\eta>\ets,~r>\rs$. \cr}
\label{eq:Lk}
\enq

Internal shocks can occur when the flow has variations in the initial $\eta$ or 
$\Lo$ with consequent variations in the terminal Lorentz factors, so shells with 
different $\Gamma_s$ catch up with each other. The shocks cannot occur at $r< \rs$ 
since in this region both shells accelerate at the same rate $\Gamma \propto r$ 
and do not catch up. After
$\Gamma$ has saturated at $r> \rs$, shocks develop at radii $\sim 2 c \tv \Gamma_1 
\Gamma_2 \sim 2 c \tv \Gamma^2$ for shells whose terminal Lorentz factors differ
by $\Delta \Gamma= \Gamma_2 - \Gamma_1 \sim \Gamma$ as a result of initial 
variations in $\eta$ or $L$ over timescales $t_v = \xi_v \to < t_w$, where $\xi_v 
\geq 1$ and $\to$ is the minimum dynamic timescale in equation (\ref{eq:ro}).
\beq
(\rsh / \ro )= 2.17\times 10^5 \xi_v \m1 \Gamma_2^2 =
   2.17\times 10^1 \xi_v \ets^2 (\Gamma/\ets)^2~.
\label{eq:rsh}
\enq
(The factor $21.7=2\times 2\pi \times \sqrt{3}$ comes from the factor $2 c t_v$ in
the shock radius definition, the $2\pi$ from taking the rotation time at $\ro$ 
rather than crossing time, and $\sqrt{3}$ because $\ro$ is at three Schwarzschild 
radii.) 

To produce non-thermal radiation, shocks must occur in an optically thin region,
which requires
\beq
\eta > \eta_{sh,m}= 1.42\times 10^2 \L52^{1/5} \m1^{-1/5} Y^{1/5} \xi_v^{-1/5}
                                           ~,~ \hbox{for}~ \rsh > \rph^>,
\label{eq:etshm}
\enq
in order for a shock to occur above a photosphere which is in the coasting region,
$\rsh > \rph^> >\rs$. For $\xi_v=10^3$ corresponding to variability timescales 
$10^3 \to \sim 1$s, this can be as low as $\eta_{sh,m}\sim 35$. For even higher 
loads such that $\eta <\eta_{sh,1m}$, the photosphere is further out on the 
coasting regime, and any shocks would occur inside the photosphere.
Internal shock radii are shown in Figure 1 as a function of $\eta$ for various 
multiples $\xi_v=t_v/\to$ of the Kepler time at the last stable orbit $\to$.
However, from causality considerations shocks may occur at even lower radii, 
formally corresponding to $\xi_v \geq 1/21.7$, since as soon as the $r>\rs$ 
coasting regime is reached shells
of different $\eta$ can catch up with each other. This would allow smaller
regions where the variability timescale is as small as $\ro/c$. 

For very low baryon loads or very high $\eta >\ets$, the photosphere arises in 
the accelerating region, and in this region shocks are not possible. 
They are, however, possible beyond $\rs = r_\ast= 
\ro \ets \Go^{-1}$, where the baryon Lorentz factor has saturated to the
value $\Gamma_s=\ets$. Any initial variations on a timescale $\tv$ will lead to 
shocks at a radius $\rsh$ given by equation (\ref{eq:rsh}) with $\Gamma_s \sim 
\ets$, which will be located above $\rs=r_\ast$. There is thus a maximum possible 
internal shock radius for any given $\xi_v$, 
\beq
r_{sh,M}= 2.17 \times 10^1 \xi_v \ets^2 \ro = 
          2.17\times 10^{14} (\L52 \m1 \Go Y)^{1/2} \xi_v ~\hbox{cm}~,
\enq
which, unless $\xi_v$ is large or the external density is very large, can still 
be smaller than the radius where external shocks are expected. The range
where internal shocks can occur is shown as the horizontally or vertically 
striped region in Figure 1.

The internal shocks in the wind can dissipate a fraction of the terminal kinetic 
energy luminosity $\Lk$ above the saturation radius $\rs$. For a mechanical 
efficiency $\epsh=10^{-1}\esh1$ of conversion of kinetic energy $\Lk$ into random 
energy which can be radiated, the shock luminosity in the radiative regime is 
\beq
{\Lsh  \over \Lo } = \cases{ 
            10^{-1} \esh1 ~,                &~ for $\eta<\ets,~r>\rs$; \cr
            10^{-1} \esh1 (\eta/\ets)^{-1} ~,   &~ for $\eta>\ets,~r> \rs$. \cr}
\label{eq:Lsh}
\enq

\section{ Photosphere and Shock Spectra: Comptonization and Pair Formation}

The basic photospheric spectrum is that of a blackbody, $x F_x \propto 
x^3 \exp(-x/\Thph)$, with a thermal peak at
\beq
 x_{ph} \sim 3\Thph \leq x_{pho}= 3\Tho~, 
\label{eq:xph}
\enq
where $\Thph,\Tho$ are given by equations (\ref{eq:Thph},\ref{eq:Tho}). Notice 
that at $\rs$ the level is $\Lpht=\Lo$ and the spectral peak is at $x_{ph}\sim 
3\Theta_o\sim 1$, from equations (\ref{eq:Lpht}) (\ref{eq:Thph}). For $\rph >\rs$, 
both $\Lpht$ and $\Thph$ decrease $\propto (\rph/\rs)^{-2/3}$. 
In an $xF_x$ or $x L_x$ diagram the thermal peak $x_{ph}$ (labeled with T in
Figure 2) moves down and to the left with a slope 1 as $\eta$ decreases.
Before leaving the photosphere, however, the blackbody photons can act as seeds 
for scattering to higher energies, if there is a substantial amount of energy in 
scattering centers moving with characteristic speeds or energies larger than that 
of the emitting electrons (which are subrelativistic since $\Thph' \leq \Thp' = 
0.03$). Alfv\'en waves generated by magnetic field reconnection or MHD turbulence
can act as such centers (Thompson 1994). 
Alfv\'en waves travel at speeds $V_w$ nearly the speed of light, with an
equivalent comoving electron energy $\Theta'_w= k T'_w/m_e c^2=(1/3) 
\langle \gamma_w^2-1 \rangle \simeq (1/3)\langle V_w^2/c^2 \rangle \siml (1/3)$, 
where $V_w/c \siml 1$, and these waves can be efficiently damped for $\tau_s > 1$.
Alternatively, shocks which occur inside the photosphere may also induce Alfv\'en 
waves.  Repeated scattering on the Alfv\'en waves acts in the same way as 
comptonization off hot electrons. The spectrum follows from conservation of photon 
number and conservation of energy. As seen in the observer frame, starting from 
seed photons at energy $\Thph \leq \Tho$ this yields a spectrum $F_{x} \propto 
x^0$ or $xF_{x} \propto x$, as the conserved photon number is scattered up in 
energy.  The photons diffuse up and to the right on a slope $\propto x$ in the 
$xF_x$ diagram. From conservation of energy, the maximum energy they can reach is 
$\sim 3\Tho\sim 1$. However, if the energy in Alfv\'en waves or turbulence is a 
fraction $\eps_w < 1$ of the total $\Lo$, the comptonized photosphere luminosity is
\beq
L_{phc} =\eps_w \Lo~,
\label{eq:Lphc}
\enq
and the comptonized photosphere spectrum $xF_x \propto x$ can extend only up to 
a break energy 
\beq
x_{phc}= \min[\eps_w 3\Tho , (L_{phc}/L_{ph})3\Thph] \leq 3\Tho ~.
\label{eq:xbr}
\enq
This is still much less than the wave equivalent energy in the lab-frame, 
$\Theta_w =(k T'_w/m_e c^2)\eta \siml (1/3)\Gamma$. 
Thus above this break energy, an increasingly smaller fraction of 
the total photons can be scattered with spectrum $x F_x \propto x^0$ up to 
a maximum energy in the lab frame
\beq
x_w \sim 3\Theta_w \siml \Gamma ~.
\label{eq:xw}
\enq
As discussed by Thompson (1994) and in classical references on comptonization,
such a spectrum $xF_x \propto x^0$ is naturally expected from the diffusion of
photons out of bounded scattering regions (such as reconnection hot-spots
or turbulent cells in this case). We show in Figure 2 the comptonized photosphere
component (labeled PHC), assuming a turbulent wave energy level $\eps_w=10^{-1}
\Lo$, for various values of $\eta$, while in Figure 3 a lower value 
$\eps_w=10^{-2}$ is assumed. 

Internal shocks outside the photosphere, expected in the coasting regime if the 
outflow is unsteady, provide a significant nonthermal component of the spectrum.  The 
primary energy loss mechanism in the shocks is synchrotron radiation, or inverse 
Compton (IC). It is common to assume that at the shocks the magnetic field energy 
density is some fraction $\epb$ of the equipartition value with the outflow. The
dimensionless field at the base of the flow is then
\beq
x_{Bo}=(B/B_Q)=(2\Lo \epb/\ro^2 c)^{1/2} B_Q^{-1}=2 \L52^{1/2}\epb^{1/2}\m1^{-1}~,
\label{eq:xbo}
\enq
where $B_Q= 2\pi m_2^2 c^3/eh =4.44\times 10^{13}G$ is the critical field.
Thus the dimensionless comoving field at the shock is
\beq
x'_{Bsh}=x_{Bo}(\rsh/\ro)^{-1}\Gamma^{-1}=
    10^{-7} \Gamma_2^{-3} \L52^{1/2} \epb^{1/2} \xi_v^{-1} \m1^{-1} ~.
\label{eq:xbsh}
\enq
Note that if the fields are turbulently generated and equipartition is with 
respect to $\Lsh$, instead of $\Lo$, then from equation (\ref{eq:Lsh}), 
$\epb \leq \min[\epsh, \epsh(\ets/\eta)]$.
This is a time average value of $B'$ over the duration of the shocks, and it 
neglects any time-varying compression factors associated with individual pulses.
The observer-frame dimensionless synchrotron peak frequency in units of
electron rest mass is $x_{sy}=(3/2)x'_B \gamma_m^2 \Gamma$, where the minimum
electron random Lorentz factor in internal shocks $\gamma_m \sim 0.9\times 10^3\eep$ 
is typically a fraction $\eep$ of the equipartition value $0.5 m_p/m_e$, 
remembering that internal shock collide at relative speeds $\Gamma_{rel} \sim 1$. 
Thus the observed synchrotron peak is at
\beq
x_{sy}= x_\ast (\Gamma/\ets)^{-2}=
  1.26\times 10^1 \Gamma_2^{-2} \L52^{1/2}\epb^{1/2}\eep^2 \m1^{-1}\xi_v^{-1}~,
\label{eq:xsy}
\enq
where $x_\ast= 1.26 \times 10^{-1} \epb^{1/2} \eep^2 \m1^{-1/2} (Y \Go)^{-1/2} 
\xi_v^{-1}$. The comoving synchrotron cooling time is $t'_{sy}=2.5\times 10^{-19} 
{x'}_B^{-2}\gamma^{-1} = 2.5\times 10^{-8} \Gamma_2^6 \L52^{-1}\epb^{-1}\eep^{-1}
\m1^2 \xi_v^2$ s, using equation (\ref{eq:xbsh}).
For $\epb \ll 1$ the IC mechanism can become important for the MeV radiation. 
However, for $\epb$ not too far below equipartition values, the IC losses occur 
on timescales comparable or longer than synchrotron, and produce photons well 
above the MeV range where breaks and anomalous slopes occur.
The comoving expansion time is $t'_{ex}=r/c\Gamma=0.65 \Gamma_2 \m1\xi_v$ s, and
the ratio $t'_{sy}/t'_{ex}$ exceeds unity only for rather large $\Gamma \simg 
3\times 10^3 \L52^{1/5}\epb^{1/5}\eep^{1/5}\m1^{-1/5}\xi_v^{-1/5}$ if 
$\gamma \geq \gamma_m$. Thus the electrons are in the radiative regime above the 
synchrotron break $x_{sy}$ in all cases considered, and also for a large range 
below it.  Above the break $x_{sy}$ the synchrotron
spectrum is then $x F_x \propto x^{0}$, or more generally $x F_x \propto 
x^{-(p-2)/2}$, from the power law electrons $N(\gamma)\propto \gamma^{-p}$ 
above $\gamma=0.9\times 10^3 \eep$ produced by Fermi acceleration in the shocks
(for the rest of the discussion we assume $p=2$ or $\propto x^0$ above the break
as an example).  Below $x_{sy}$ one has $x F_x \propto x^{1/2}$, down to a 
synchrotron self-absorption frequency 
\beq
x_a \leq 10^{-4} \Gamma_2^{-4/5} \L52^{3/10}\esh1^{2/5}\eps_{B-1}^{-1/10}
   \eep^{-4/5}\m1^{-3/5}\xi_v^{-3/5}~,
\label{eq:xa}
\enq
where the equality applies if $xF_x\propto x^{1/2}$ down to $x_a$, and the
inequality applies if there is a transition to an adiabatic regime $\propto x^{4/3}$
before reaching $x_a$ (for very high $\Gamma$ or $\xi_v$).
Schematic synchrotron spectra are shown in Figure 2 from shocks where 
$\epsh=10^{-1}$, and in Figure 3 (left panel) where $\epsh=3\times 10^{-3}$,
assuming $p=2$. These curves, labeled S, show the break at $x_{sy}$ and a
radiative slope 1/2  below that down to $x_a$. For cases where an adiabatic 
regime is achieved at frequencies below $x_{sy}$ but above $x_a$, the synchrotron 
slope would steepen to 4/3 and $x_a$ moves further down in frequency.

Pair breakdown via $\gamma\gamma \rightarrow \epm$ can occur when the comoving 
synchrotron luminosity from the internal shocks is large enough, which is the case
over a non-negligible range of parameter space.  The comoving
compactness parameter is $\cl =n'_\gamma \st (r/\Gamma)$, where $n'_\gamma
=(\alpha \Lsh /4 \pi \rsh^2 m_e c^3 \Gamma^2)$ is the comoving photon density 
and $\alpha$ is the fraction above threshold. For typical synchrotron spectra 
peaking at $x_{sy}$ and with $x F_x \propto x^0$ above that, a fraction above 
threshold in the comoving frame $\alpha=0.3 \alpha_{-.3}$ is a typical value 
for the high $\cl$ cases. 
Thus
\beq
\cl=(\alpha \Lsh \st / 8\pi m_e c^3 t_v \Gamma^5) \simeq
    3\times 10^2 \Gamma_2^{-5}\L52 \esh1 \alpha_{-.3} \xi_v^{-1} ~.
\label{eq:compac}
\enq
For values of $\Gamma> \Gamma_{\cl} \sim 3.1 \times 10^2 (\L52 \esh1 \alpha_{-.3} 
\m1^{-1} \xi_v^{-1})^{1/5}$ the compactness $\cl \leq 1$ and pair formation does 
not occur. This corresponds to shock radii $(\rsh/\ro)_{\cl} \geq 0.7\times 
10^8 \L52\esh1\alpha_{-.3}\m1^{-1} \Gamma_2^{-3}$.  Below that, in the range 
$\rs < \rsh < r_{sh\cl}$, one has $\cl \simg 1$, and pair breakdown rapidly adds 
to the opacity and cooling. This leads to a self-regulating pair plasma whose 
scattering optical depth $\tau_s \sim$ several, with a characteristic mean $\epm$ 
energy $\Theta'_{c} \sim m_e c^2/\tau_s \sim 10^{-1}$ (e.g. Svensson 1987, 
Ghisellini \& Celotti 1999). 
Clearly, $\tau_s$ cannot be too large, otherwise advection and adiabatic cooling
would dominate over comptonization and diffusion. At such nonrelativistic 
energies, cyclotron radiation produces seed photons at harmonics whose energy is 
much lower than $x_{sy}$, but repeated scattering on the much hotter $\epm$ 
produces a comptonized spectrum $F_{x'} \propto {x'}^0$ up to the maximum energy 
$\Theta'_{c}$. For $\Gamma =\eta$ (and $\cl >1$), one then has an observer-frame
characteristic break energy
\beq
x_{c} \sim 10^{-1} \Gamma \sim 10\Gamma_2 ~,
\label{eq:xic}
\enq
and below that a spectrum $xF_x \propto x$. The pair Comptonized luminosity is 
comparable to the (initial) synchrotron shock luminosity in the absence of pairs, 
$L_{c}/\Lo= \Lsh/\Lo$ given by equation (\ref{eq:Lsh}), since the self-regulation 
of the comptonizing pair plasma is achieved by a time-averaged balance between 
energy dissipation by the shocks and radiative losses. 

Above the break $x_{c}$, one would expect a drop-off of the spectrum, in the 
absence of other effects.  However, since internal shocks occur in 
the coasting regime and self-regulating pair breakdown tends to maintain a 
moderate scattering optical depth, reconnection and MHD turbulence may arise
here too, leading to Alfv\'en waves of much higher characteristic energies
than that of the pairs, which could lead to a flatter power law spectrum 
$x F_x \propto x^0$ extending above $x_{c}$ up to energies $x_w$ similar to 
that given in equation (\ref{eq:xw}).
In Figure 2 the top row shows cases where $\cl <1$ and hence only the photosphere
(with thermal T and nonthermal PHC) and the shock synchrotron (S) components
are present. The middle row shows the marginal cases where $\cl=1$ and hence
besides the photosphere T and PHC components one expects the synchrotron S and the 
pair breakdown comptonized C components to be comparably important. The bottom row 
of Figure 2 shows cases with $\cl \gg 1$, where pair breakdown is so important as 
to completely replace the synchrotron component S with a self-regulated
comptonizing pair plasma component C. Both the S and C components have a luminosity 
level given by a shock efficiency (e.g. Kumar 1999) $\epsh=10^{-1}$ in Figure 2, 
with the effects of a lower value $\epsh=3\times 10^{-3}$ shown in Figure 3 (left).

\section{Discussion}

Within the framework of the standard internal shock model, 
we have analyzed the observable effects of the two major radiating regions,
the photosphere and the internal shocks, which are expected to contribute to the 
flux from an unsteady fireball outflow model of GRB. Note that, as in most GRB
models, the spherically symmetric assumption applies to any conical outflow
whose opening half-angle is $\simg \Gamma^{-1}$. (We note also that, if the
cone angle {\it were} very narrow, transverse pressure gradients would cause
significant departures from radial outflow; under these circumstances, the 
$r$-dependences would change, although the qualitative features would not 
be substantially altered.)
We have purposely left out of our discussion any radiation from an external 
shock, which is expected to occur at radii beyond those considered here (and 
which can add other radiation components, especially a long term afterglow).

The standard internal shock model of GRB is generally assumed to produce its
observed nonthermal radiation by the synchrotron (or possibly inverse Compton)
process. Here, in addition to synchrotron we have also considered in more detail 
the role of the outflow photosphere and of possible nonthermal spectral 
distortions in it, as well as the role of pair breakdown in shocks with very 
high comoving luminosity. In a diagram (Figure 1) of radius vs. dimensionless 
entropy $\eta={\dot L}/{\dot M} c^2$,
the regions where the internal shocks are dominated by synchrotron radiation 
(and pair breakdown is unimportant, $\cl \leq 1$) are shown by the vertically 
striped area S. The line where the compactness parameter $\cl=1$ runs parallel 
to the line $\rph^>$ for the photosphere in the coasting regime, in the same
figure. The region where internal shocks are dominated by pair-breakdown, $\cl >$,
is given by the horizontally striped region C in Figure 1, where the shock 
spectrum is dominated by comptonizing pairs. 
To the left and below the photospheric lines, shocks would occur at high optical 
depths and their spectrum would be thermalized, adding to the purely thermal 
(T) and non-thermal (PHC) components emerging form the photosphere. These various 
spectral components are shown in Figures 2 and 3 in a power per decade 
$x L_x$ vs. $x$ plot, where $x$ is photon energy in electron rest mass units.

In our earlier papers on fireball shock models of GRB
and most subsequent related work, the role of photospheres 
and pair breakdown was briefly considered, but until recently the observations
did not appear to provide much support for their being important. The need for 
a non-thermal spectrum continues to be a strong argument for shocks and a 
synchrotron component, while for the less problematic case of large baryon loads 
the photospheres occur in the coasting regime, where their observer frame thermal
luminosity is drastically weakened by adiabatic cooling. This is confirmed by the 
spectra of Figures 2 and 3, where 
for moderate to large baryon loads (low $\eta$) and moderate variability $\xi_v 
\geq 10^2$ or $t_v \simg 0.3$ s the thermal peaks T are strongly suppressed, 
especially in the more ``conservative" region going part way above and below from 
the central and right-of-center panels of Figure 2.  
Pair breakdown is also a phenomenon considered in earlier papers (c.f. also Pilla
\& Loeb 1998, Papathanassiou \& \Mesz 1996), which received
less attention than it deserves because its importance appears to be restricted
to a relatively narrow region of parameter space. This is illustrated in Figure 1, 
where one can compare the vertically striped region $\cl <1$ labeled S versus 
the narrowish horizontally striped region $\cl >1$ labeled C. This region would be 
even narrower if one normalized to $\Lo \sim 10^{50} \ergs$ as opposed to $\L52$.

However, fresh motivation for reconsidering the role of photospheres and pairs 
is provided by the evidence, in a non-negligible fraction of bursts, of low energy 
(1-10 keV) spectral slopes steeper than 1/3 in energy (or 4/3 in $\log xL_x$)
and in some cases an X-ray excess above the power law extrapolation from higher 
energies, as well as the ubiquitousness of observed break energies clustering 
between 50-1000 keV (discussed in \S 1 and references there). A look at Figures 2 
and 3 shows that allowing a larger role to photospheres and comptonized pairs 
provides a way of addressing these observational trends. If photospheres are rather 
more important than shocks (e.g. due to some preference for very high $\eta$, 
weakly varying outflows or low shock efficiencies) the thermal component can 
provide low energy slopes as steep as $\propto x^3$ in $xL_x$ (while flatter slopes 
can be achieved through integration or distributions). The same can
explain an X-ray excess well above the power law extrapolation from above.
Reasonable break energies can be obtained from either the synchrotron or pair 
comptonization mechanisms in the shock, but they depend on $\Gamma=\eta$, and 
unless the range of $\eta$ is narrow they would not necessarily cluster between
50-1000 keV. In the case of pair comptonization they also tend to be a bit high, 
unless the equilibrium pair temperature is $\Theta'_c \siml 0.03$.
A preferred break could be attributed  to a photospheric peak, provided 
baryon loads are very low, $\eta \geq \ets \sim 10^3$ in all cases; this still 
requires a strong shock synchrotron component, or possibly Alfv\'en wave 
comptonization in the photosphere, to explain the high energy power law spectra.
It would also imply a pronounced upward change of slope above the break, from 
the thermal peak to a flatter power law in all bursts where a break is observed, 
unless the shock or Alfv\'en wave scattering always produces a luminosity 
comparable to the thermal photosphere. Preferred break energies arise naturally 
if Thompson's (1994) proposed mechanism of comptonization by Alfv\'en wave damping 
in the photosphere is taken at face value and has good efficiency, leading to a 
source frame break around 0.5 MeV. 
Note that the comptonized spectral slopes discussed here (either from thermal pairs 
in shocks or Alfv\'en waves in the photosphere) are nominally 1 and 0 in $xF_x$,
but for simple time-dependent calculations an evolving slope is expected (e.g. 
Liang \etal, 1999, Thompson 1994). In reality the actual time dependence for an 
unsteady outflow leading to shocks, pair breakdown and comptonization could be 
more complicated.

If both a photospheric and a shock component are detected, one would expect the
thermal photospheric luminosity (and its non-thermal part, if present) to vary on
similar timescales as the nonthermal synchrotron or pair comptonized shock
component (unless the shock efficiency is radius dependent, or unless one or both 
are beyond $r=\ro \eta^2$, in which case $r/c\eta^2$ imposes a lower limit on the 
corresponding variability timescale). But even if the bolometric luminosity 
varies on the same timescale, the luminosity in a given band (e.g. BATSE) 
probably varies differently, since the thermal peak energy is $\propto L^{1/4}$
and falls off steeply on either side, while the synchrotron peak energy varies 
$\propto B'\eta \propto L^{3/2}$ and falls off more slowly. The pair comptonized
break energy on the other hand varies as $\eta\propto L$, and also drops off
slowly on the low energy side, or on both sides if scattering off waves is
present in the high energy side.
If observations at high time resolution become possible in X-ray or optical
during the burst (as opposed to the afterglow), we would expect (c.f. Figure 2)
the bursts with shorter time structure (low $\xi_v$) to be more suppressed at 
these wavelengths, compared to those with longer variability timescales.

In summary, there are several plausible mechanisms for producing a preferred 
energy break, which rely on internal properties of the outflow. In those bursts
with low energy slopes steeper than implied by synchrotron, the prominence of the 
photosphere is a likely explanation, in which case its luminosity could in some
cases vary differently from the higher energy power law component. Intrinsically
high luminosity bursts, where pair breakdown is inferred, would be predicted to
have generally harder high energy power slopes than in lower luminosity bursts
where synchrotron provides the high energy slope. If photospheric comptonization
on Alfv\'en waves is responsible for the high energy power law slopes, the thermal 
peak and the power law should vary together in time. In this case a 
straightforward prediction is that the thermal peak photon energy and the 
intrinsic luminosity allow one to determine the expected maximum photon energy 
$x_w\sim \Gamma$.

\acknowledgements{This research has been supported by NASA NAG5-2857, 
NSF PHY94-07194 and the Royal Society. }

\newpage

\begin{figure}[htb]
\centering
\epsfig{figure=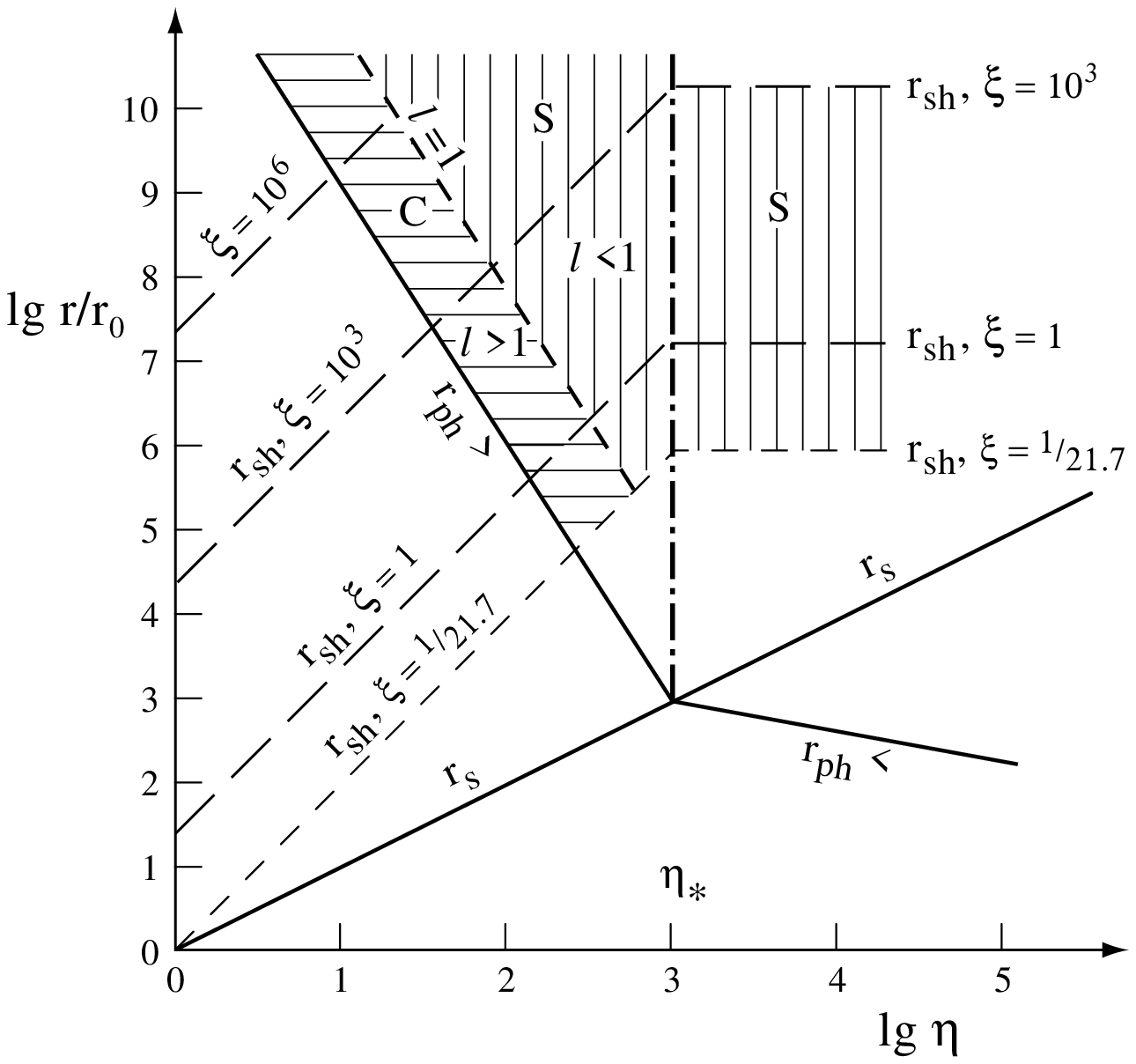, width=6.in, height=6.in}
\figcaption[eta_r.eps]{
Radius vs. $\eta=L/{\dot M}c^2$  diagram showing the location of the
photosphere $\rph$, the saturation radius $r_s$, and the shock radii $\rsh$ for
various variability timescales $\xi_v$. The regions where the shock spectrum is
dominated by synchrotron (S) or comptonization in a pair plasma (C) correspond
to a comoving compactness $\cl<1$ or $\cl>1$.
   \label{fig:radius}
}
\end{figure}

\begin{figure}[htb]
\centering
\epsfig{figure=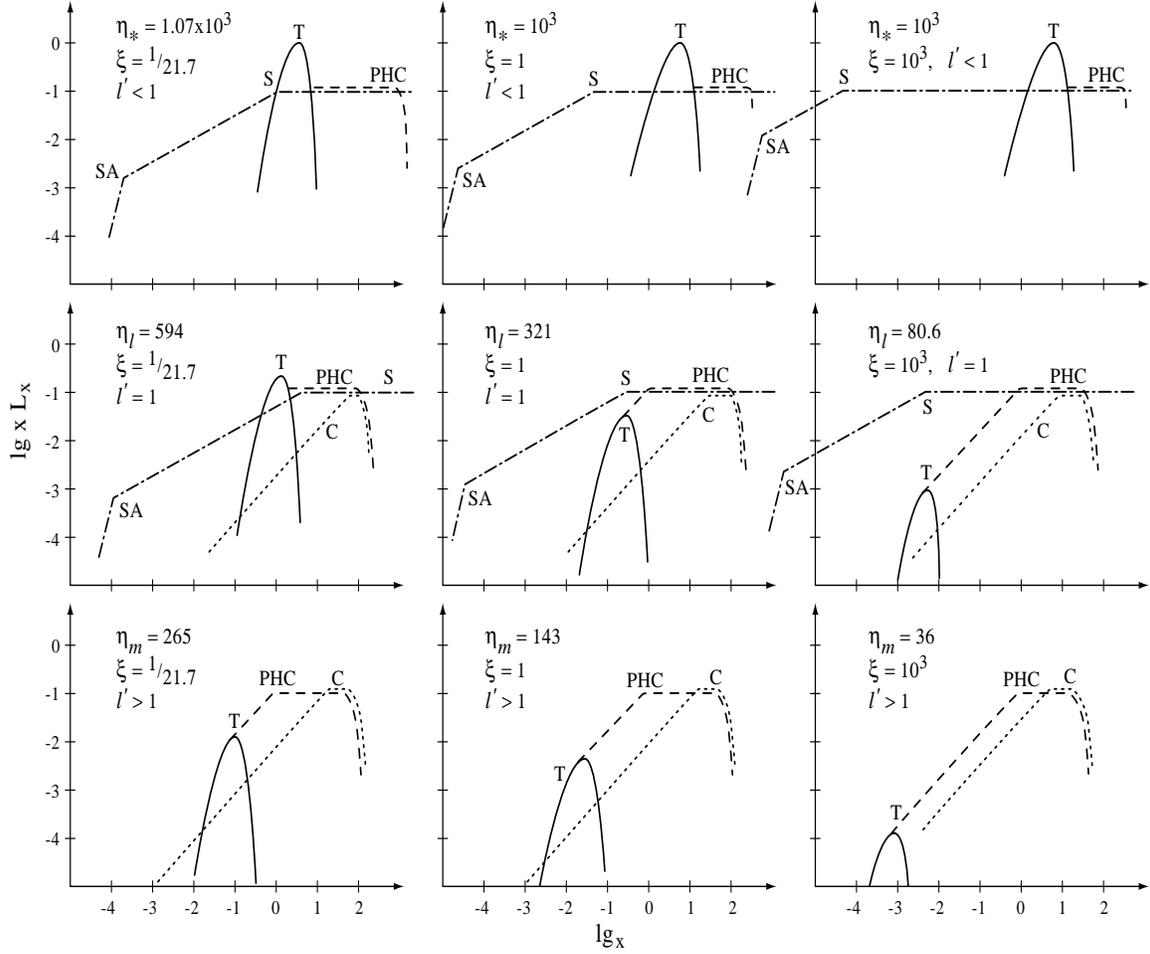, width=6.in, height=5.in}
\figcaption[eta_xi1.eps]{
Luminosity per decade $xL_x$ vs. $x=h\nu/m_e c^2$ for different values of
$\eta$ increasing upwards, and different variability timescales increasing
to the right, using $\eps_w=\epsh=10^{-1}$. Pair formation is unimportant in the
top row and important in the bottom row, being marginal in the middle row.
T: thermal photosphere, PHC: photospheric comptonized component; S: shock
synchrotron; C: shock pair dominated comptonized component.
   \label{fig:spec6}
}
\end{figure}

\begin{figure}[htb]
\centering
\epsfig{figure=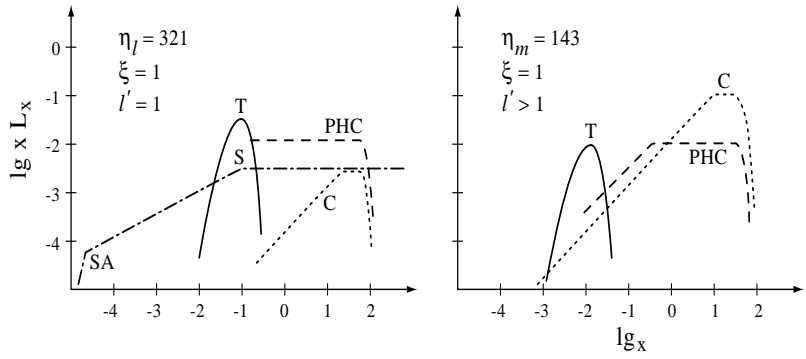, width=4.2in, height=1.85in}
\figcaption[eta_xi2.eps]{
Luminosity per decade $xL_x$ vs. $x$ for two of the spectra in
Figure 2 but with different shock synchrotron and pair comptonized components
or lower comptonized photosphere component. Left panel: $\eps_w=10^{-2},\epsh=
3\times 10^{-3}$. Right panel: $\eps_w=10^{-2},\epsh=10^{-1}$.
   \label{fig:spec2}
}
\end{figure}

\end{document}